\documentclass[epj,final,numComoabook]{webofc} 
\usepackage[varg]{txfonts}   
\usepackage{subfigure}
\newcommand{\dAu}{\mbox{${\rm {\it d}+Au}$}\xspace}

\newcommand{\HeAu}{\mbox{${\rm He+Au}$}\xspace}
\newcommand{\AuAu}{\mbox{${\rm Au+Au}$}\xspace}
\newcommand{\UU}{\mbox{${U+U}$}\xspace}

\newcommand{\CuCu}{\mbox{${\rm Cu+Cu}$}\xspace}
\newcommand{\CuAu}{\mbox{${\rm Cu+Au}$}\xspace}
\newcommand{\pp}{\mbox{${p+p}$}\xspace}
\newcommand{\AAA}{\mbox{${\rm A+A}$}\xspace}

\newcommand{\eHF}{\mbox{${e^{\pm}_{HF}}$}\xspace}

\newcommand{\mmHF}{\mbox{${\mu^{-}_{HF}}$}\xspace}
\newcommand{\Jpsi}{\mbox{${J/\psi}$}\xspace}
\newcommand{\po}{\mbox{$\pi^{0}$}\xspace}

\newcommand{\pT}{\mbox{${p_{\rm{_{T}}}}$}\xspace}

\newcommand{\Ncoll}{\mbox{$N_{\rm coll}$}\xspace}

\newcommand{\snn}{\mbox{$\sqrt{s_{_{NN}}}$}\xspace}

\newcommand{\Np}{\mbox{${\rm N_{part}}$}\xspace}

\newcommand{\bim}{\mbox{${b}$}\xspace}

\newcommand{\RAA}{\mbox{${R_{AA}(\pT)}$}\xspace}
\newcommand{\RdAu}{\mbox{${R_{dAu}(\pT)}$}\xspace}

\def\FigureOne{
\begin{figure}[h]
\centering
    \subfigure[\eHF and \po : \AuAu vs. \dAu at \snn = 200 GeV]
    {
        \includegraphics[width=0.45\textwidth]{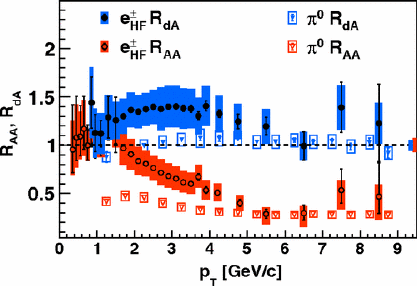}
        \label{fig:fig1a}
    }
    \subfigure[\eHF and \po : \CuCu at \snn =  62.4 GeV]
    {
        \includegraphics[width=0.45\textwidth]{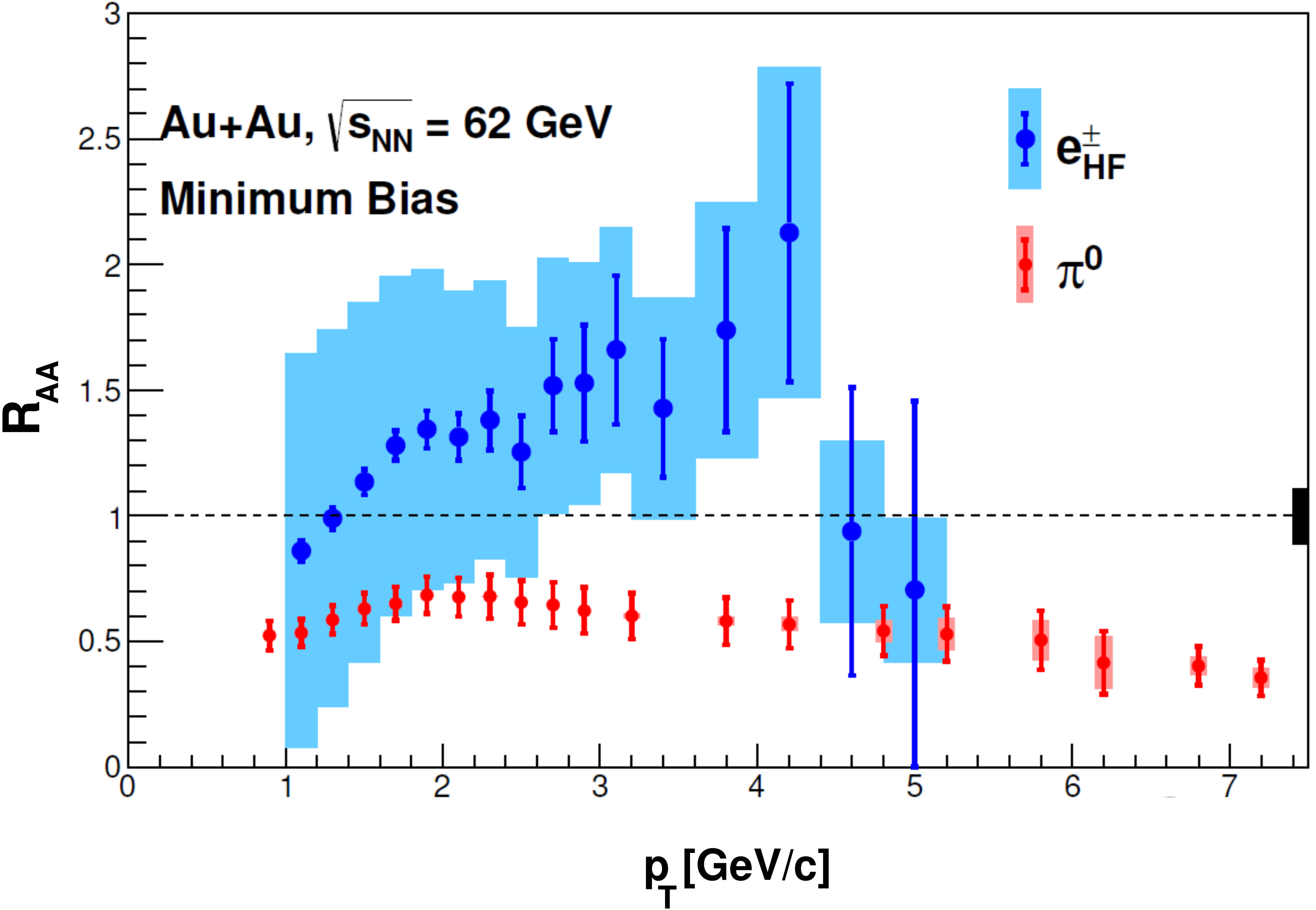}
        \label{fig:fig1b}
    }
    \\
    \subfigure[\eHF and \mmHF : \CuCu at \snn =  62.4 GeV]
    {
        \resizebox{0.5\textwidth}{!}{\includegraphics{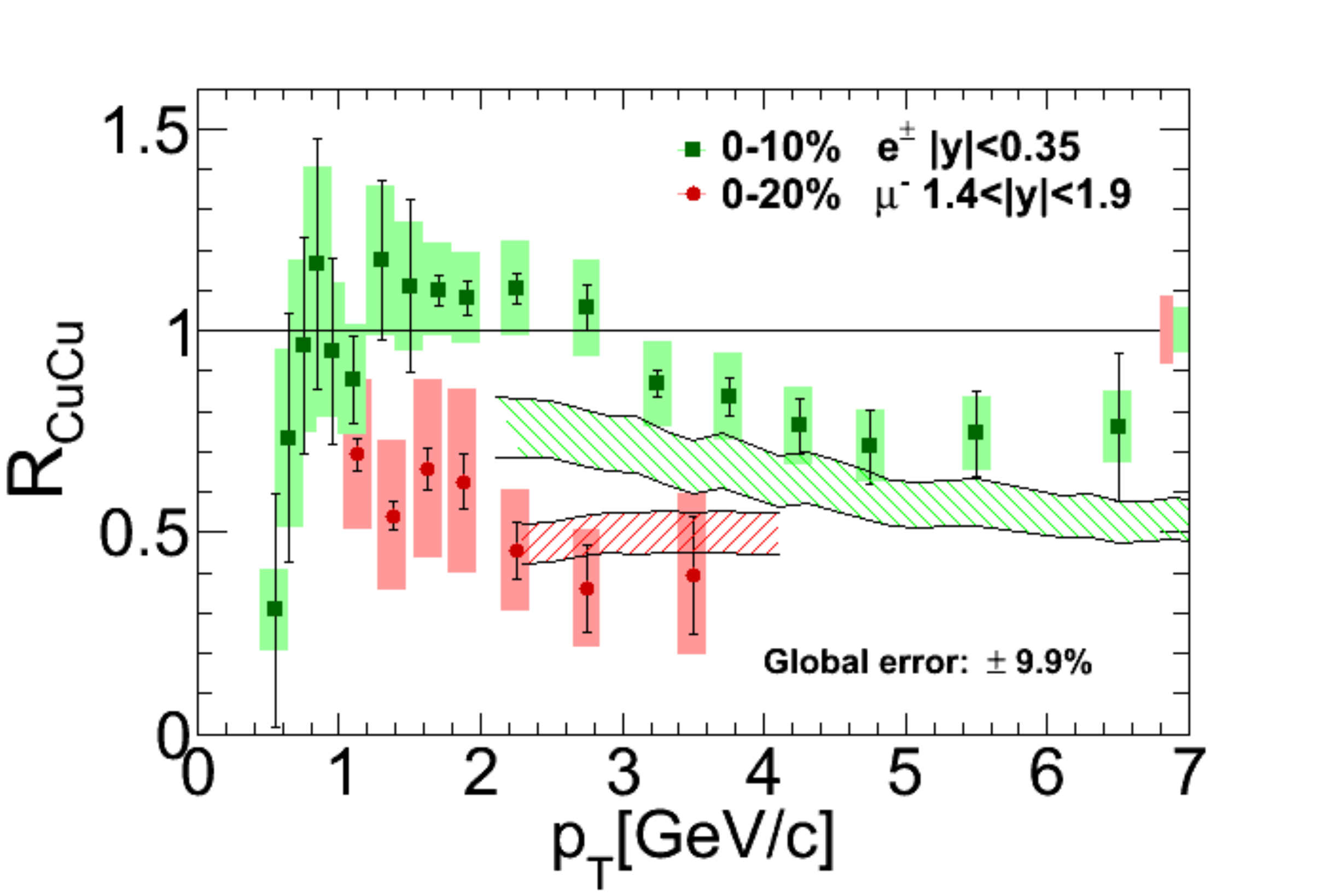}}
        \label{fig:fig1c}
    }
    \vspace*{-0.3cm}
    \caption{The nuclear modification factors \RdAu and \RAA for the
      \po, \eHF and \mmHF : panels (a) and (b) for minimum bias \dAu,
      \AuAu and panel (c) for central collisions \CuCu. The two boxes
      on the right side of the plot represent the global uncertainties
      in the \dAu, \AuAu (\CuCu) values of \Ncoll
      \cite{PHE06,PHE09,PHE12A,PHECuCuEmu,PHE07,PHEAuAu,RHICNouicer,PHEdAu,NRNouicerNN2012}. \label{fig:fig1}}
\end{figure}
}
\def\FigureTwo{
\begin{figure}[!]
\begin{minipage}{.47\textwidth}
\includegraphics[width=6.5cm ,height=7cm]{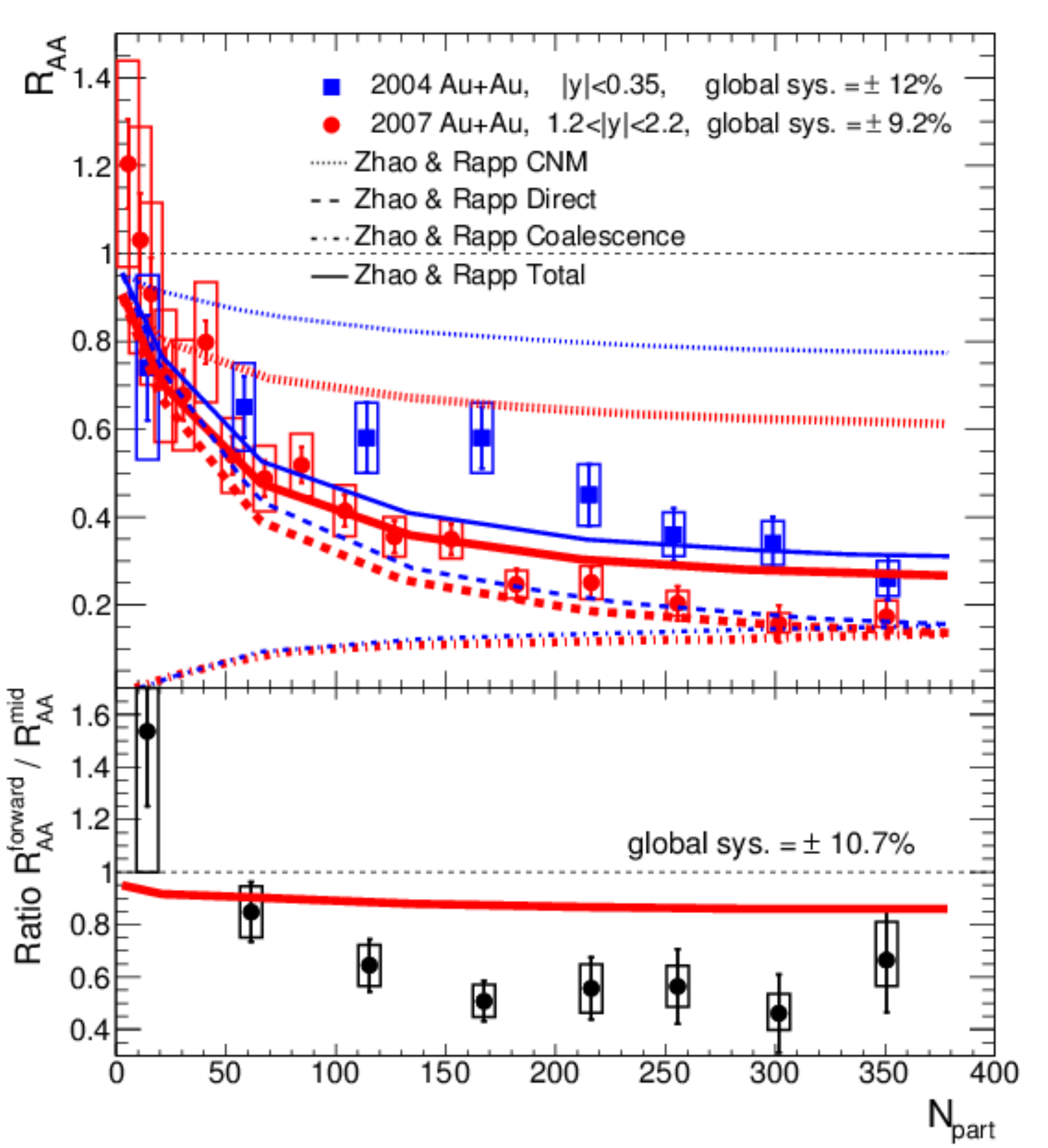}
\vspace*{-0.2cm}
\caption{\small The upper panel shows the nuclear modification factor,
  $R_{AA}$ as a function of centrality.  The lines show theoretical
  calculations: the effect of cold nuclear matter (dotted, uppermost
  pair), direct production (dashed), and coalescence (dot-dashed).
  The solid lines represent the sum of all effects considered.  The
  lower panel shows the ratio of forward- to mid-rapidity $R_{AA}$
  Ref.~\cite{PHE11A}.
\label{fig:AuAuRAA}}
\end{minipage}
\hspace*{0.1cm}
\begin{minipage}{.47\textwidth}
\includegraphics[width=6.8cm, height=4cm]{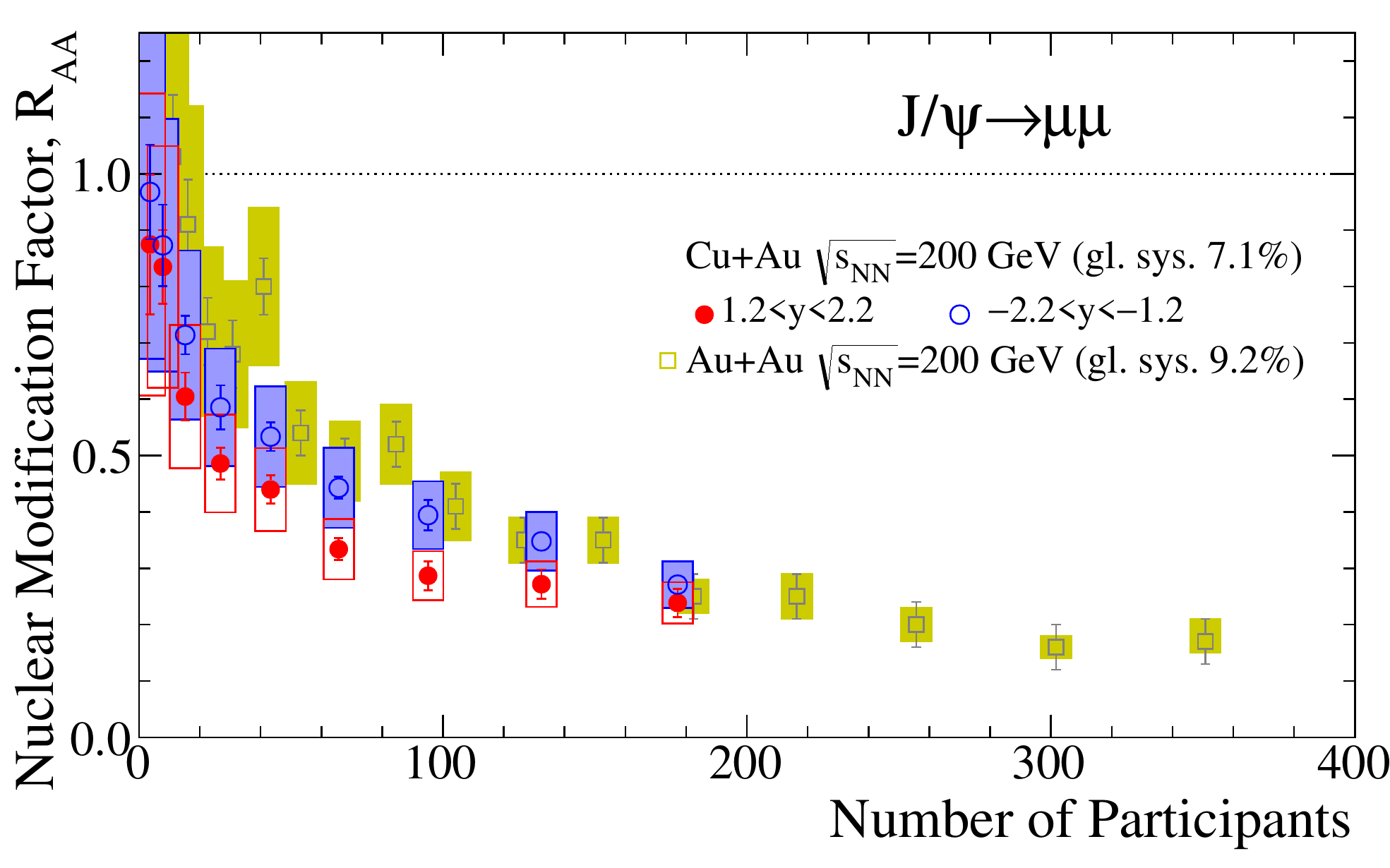}
\vspace*{-0.3cm}
\caption{\small Nuclear modification factor, $R_{AA}$ as a function of
  centrality in Cu+Au collisions at forward (red) and backward (blue)
  rapidity.  For comparison, the Au+Au (green) data are also shown,
  averaged over forward and backward rapidities
  Ref.~\cite{PHENIXCuAu}. \label{fig:CuAuRAA}}
\end{minipage}
\end{figure}
}
\def\FigureThree{
\begin{figure}[!]
\centering
\includegraphics[width=0.499\textwidth]{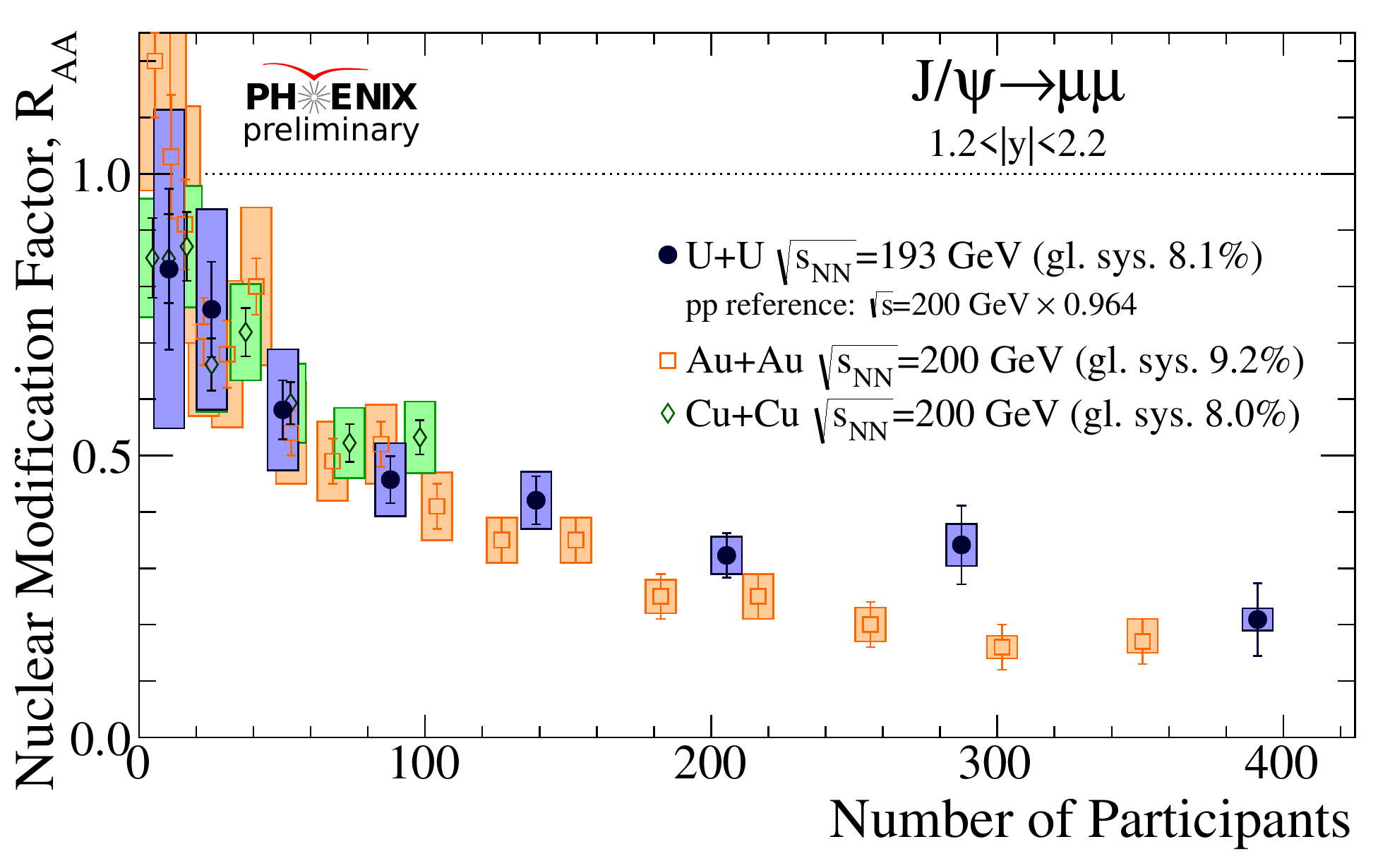}
\vspace*{-0.3cm}
\caption{Nuclear modification factor, \RAA, measured in \UU (blue) collisions
  at \snn = 193 GeV as a function of centrality.  For comparison, Au+Au (red) and
  Cu+Cu (green) data are also shown. The \UU  \RAA results are preliminary.}
\label{fig:UU}
\end{figure}
}

\begin{document}
\title{Probing Properties of Hot and Dense QCD Matter with\\ Heavy
  Flavor in the PHENIX Experiment at RHIC}
\author{Rachid Nouicer\inst{1}, for the PHENIX Collaboration}
\institute{Department of Physics, Brookhaven National Laboratory,
  Upton, NY 11973, United States}
\abstract{Hadrons carrying heavy quarks, i.e. charm or bottom, are
  important probes of the hot and dense medium created in relativistic
  heavy ion collisions. Heavy quark-antiquark pairs are mainly
  produced in initial hard scattering processes of partons. While some
  of the produced pairs form bound quarkonia, the vast majority
  hadronize into particles carrying open heavy flavor. Heavy quark
  production has been studied by the PHENIX experiment at RHIC via
  measurements of single leptons from semi-leptonic decays in both the
  electron channel at mid-rapidity and in the muon channel at forward
  rapidity.  A large suppression and azimuthal anisotropy of single
  electrons have been observed in \AuAu collisions at 200 GeV. These
  results suggest a large energy loss and flow of heavy quarks in the
  hot, dense matter. The PHENIX experiment has also measured \Jpsi
  production at 200 GeV in \pp, \dAu, \CuCu and \AuAu collisions, both
  at mid- and forward-rapidities, and additionally \CuAu and \UU at
  forward-rapidities. In the most energetic collisions, more
  suppression is observed at forward rapidity than at central
  rapidity. This can be interpreted either as a sign of quark
  recombination, or as a hint of additional cold nuclear matter
  effects.  The centrality dependence of nuclear modification factor,
  \RAA, for \Jpsi in \UU collisions at \snn = 193 GeV shows a similar
  trend to the lighter systems, \AuAu and \CuCu, at similar energy 200
  GeV. }
\maketitle
%
\section{Introduction \label{intro}}
The measurement of inclusive hadron yields in central Au+Au collisions
at RHIC led to the discovery of the suppression of hadron production
at large transverse momenta ($p_T$) compared to $p+p$
collisions~\cite{RHIC}. This is generally attributed to the energy
loss of light partons in the dense nuclear matter created at RHIC.
Heavy quarks, i.e. charm and beauty, are believed to be mostly created
in initial hard scattering processes of partons~\cite{Lin95} and thus
are excellent probes of the hot and dense matter formed in
nucleus$-$nucleus collisions at high energy. While some of the
produced pairs form bound quarkonia, the vast majority hadronize into
hadrons carrying open heavy flavor. They interact with the medium and
are expected to be sensitive to its energy density through the
mechanism of parton energy loss. Due to the large mass of heavy quarks
the suppression of small angle gluon radiation should reduce their
energy loss, and consequently any suppression of heavy-quark mesons
like $D$ and $B$ mesons at high-$p_T$ is expected to be smaller than
that observed for hadrons consisting of light quarks~\cite{Dok01}.

We quantify the medium effects on high-\pT\ production in
nucleus-nucleus collisions, \AAA, with the nuclear modification factor
which is defined as following:
\begin{equation}
{\rm R_{AA} (p_{T})= \frac{ 1 }{\langle N_{coll} \rangle} \times \frac{yield\ per\ A+A\ collision}{
    yield\ per\ {\it p+p}\ collision} =\frac{ 1 }{\langle N_{coll}\rangle} \times
  \frac{d^{2}N^{^{A+A}}/dy dp_{T}}{d^{2}N^{^{\it p+p}}/dy dp_{T}} }
\end{equation}
This factor reflects the deviation of measured distributions of
nucleus-nucleus, \AAA, transverse momentum, at given impact parameter
\bim, from measured distributions of an incoherent superposition of
nucleon-nucleon (\pp) transverse momentum, scaled by the average
number of expected binary collisions $\langle$\Ncoll$\rangle$. This
normalization often is known as ``binary collisions scaling". In the
absence of any modifications due to the `embedding' of elementary
collisions in a nuclear collision, we expect ${\rm R_{AA}} = 1$ at
high-\pT. At low \pT, where particle production follows a scaling with
the number of participants, the above definition of ${\rm R_{AA}}$
leads to ${\rm R_{AA}}$ $<$ 1 for \mbox{\pT\ $<$ 2 GeV/c}.

However, the interpretation of heavy-ion collision data is complicated
by the fact that heavy quark production is modified in a nuclear
target by cold nuclear matter (CNM) effects. These include
modification of the parton density functions in a nucleus, parton
energy loss in CMN, and (for quarkonia states), breakup due to
collisions with nucleons in the target. To disentangle such effects,
multiple measurements are needed, such as collisions of different ions
and at different collision energies.  RHIC has the unique ability to
store and collide nuclei of different mass, i.e. \AuAu, \CuCu and \UU,
and also collide asymmetric systems such as \dAu, \HeAu and \CuAu.  By
observing the modification in more elementary systems like in \dAu and
\HeAu, one may hope to understand the initial modifications before the
QGP is formed.

Heavy quark production has been studied by the PHENIX experiment at
RHIC via measurements of single leptons from semi-leptonic decays in
both the electron channel at mid-rapidity and in the muon channel at
forward-rapidity. In this paper I will summarize the latest PHENIX
results concerning open and closed heavy flavor production as a
function of beam energy and systems size.

\section{PHENIX Experiment \label{sec-1}}
The PHENIX detector~\cite{PHE03} comprises three separate
spectrometers in three pseudorapidity ($\eta$) ranges. Two central
arms at midrapidity cover $\mid \eta \mid <$ 0.35 and have an
azimuthal coverage ($\phi$) of $\pi$/2 rad each, while muon arms at
backward and forward rapidity cover -2.2 $< \eta <$ -1.2 and 1.2 $<
\eta <$ 2.4, respectively, with full azimuthal coverage. In the
central arms, heavy quark production was studied via measurements of
single leptons (electrons) from semi-leptonic decays. Charged particle
tracks are reconstructed using the drift chamber and pad
chambers. Electron candidates are selected by matching charged tracks
to hits in the Ring Imaging Cherenkov (RICH) counters and clusters in
the Electromagnetic Calorimeter (EMCal). At forward and backward
rapidity, the heavy quark production is measured via dimuon
decays. Muons are identified by matching tracks measured in
cathode-strip chambers, referred to as the muon tracker (MuTr), to
hits in alternating planes of Iarocci tubes and steel absorbers,
referred to as the muon identifier (MuID). Each muon arm is located
behind a thick copper and iron absorber that is meant to stop most
hadrons produced during the collisions, so that the detected muons
must penetrate 8 to 11 hadronic interaction lengths of material in
total. Beam interactions are selected with a minimum-bias (MB) trigger
requiring at least one hit in each of the two beam-beam counters
(BBCs) located at positive and negative pseudorapidity 3~$< \mid \eta
\mid <$~3.9.  \FigureOne
\section{Open heavy flavor - single electron/muon measurements \label{sec-2}}
Open heavy flavor production is measured in PHENIX through the
measurement of inclusive electrons or muons
\cite{PHENIXHFsingleE,PHENIXHFfor}. These analyses use a cocktail
method to remove fake and real electrons/muons from the data
sample. i.e. for the electron measurement, the electrons that come
from either meson decay or photon conversions are measured and
subtracted from the inclusive spectrum and the remainder is attributed
to electrons coming from the semi-leptonic decay of $D$~and~$B$
mesons. This remainder is also referred to as the non-photonic
electron (NPE) component. PHENIX has measured spectra of the single
electrons~\cite{PHE06,PHE09,PHE12A,PHECuCuEmu} and single
muons~\cite{PHECuCuEmu,PHE07} from heavy flavor in \pp and \CuCu
collisions as well single electrons from heavy flavor in \AuAu
\cite{PHEAuAu,RHICNouicer} and d$+$Au~\cite{PHEdAu,NRNouicerNN2012}
collisions.

The effects of CMN are expected to be present in the initial state of
\AAA collisions; however, this CNM enhancement is convolved with the
suppressing effects of hot nuclear matter. Figure \ref{fig:fig1a}
shows \RdAu and \RAA for \eHF and the neutral pion (\po), for which
only small CNM effects are observed between \eHF and \po for \pT $<$ 5
GeV/c. Above \pT $\approx$ 5 GeV/c, where the CNM effects on both
species, \eHF and \po, are small, their \RAA values are consistent
within uncertainties. However, in the range where CNM enhancement is
large for \eHF and small on \po, the corresponding \eHF\RAA values are
consistently above the \po values. This could suggest that the
difference in the initial state cold nuclear matter effects due to the
mass-dependent Cronin enhancement is reflected in the final state
spectra of these particles in \AuAu collisions, although alternate
explanations involving mass-dependent partonic energy loss in the hot
medium are not ruled out.

To study the interplay between initial-state effects and final-states
effects for heavy flavor (\eHF) productions, PHENIX measured the \RAA
in \AuAu collisions at low beam energy, \snn = 62.4~GeV. These \RAA
values for \eHF in \AuAu collisions at 62.4~GeV are compared to
\po\RAA at the same energy as shown in Figure~\ref{fig:fig1b}.  At
62.4 GeV, the competition, if present, favors heavy-flavor
enhancement over suppression. This is consistent with previous results
with hadrons where the Cronin enhancement increases as the collision
energy decreases \cite{PHE12A}.

PHENIX measurements of heavy flavor muons (\mmHF ) at forward rapidity
(1.4 < y < 1.9) in central Cu+Cu collisions at \snn = 62.4 GeV show
a significant suppression \cite{PHE12A}. The magnitude of this
suppression at forward-rapidity in \CuCu (shown in
Fig.\ref{fig:fig1c}) is compared to the suppression of \eHF in central
Au+Au collisions at mid-rapidity at the same energy. As open
heavy flavor is significantly more suppressed at forward-rapidity than
at mid-rapidity in \CuCu, additional nuclear effects, such as gluon
shadowing at low-Bj\"{o}rken-$x$ or partonic energy loss in the
nucleus, may be significant. The heavy flavor \eHF and \mmHF are
compared in Fig.\ref{fig:fig1c} to a theoretical prediction that
combines the effects of partonic energy loss, energy loss from
fragmentation and dissociation, and includes nuclear matter effects
such as shadowing and Cronin enhancement due to parton scattering in
the nucleus \cite{Shar2009}. While consistent within uncertainties,
the model predicts more suppression for heavy flavor electrons than
seen in the data.  \FigureTwo

\section{Heavy quarkonia - \Jpsi production versus system size \label{sec-4}}
Heavy quarkonia has long been proposed as a sensitive probe of the
color screening length and deconfinement in the quark-gluon
plasma~\cite{Mue05}. It was originally suggested by Matsui and
Satz~\cite{MatsuiSatz} that a signature of the formation of the QGP
would be that $J/\psi$ states that are initially bound may
disassociate due to color screening in the high temperature QGP
matter. The picture that was originally proposed is complicated by
competing effects, which modify quarkonia production and survival in
cold and hot nuclear matter.  Heavy quarkonia measurements are made in
PHENIX by either detecting opposite sign electrons at mid-rapidity
($\mid$y$\mid <$0.35) or by detecting opposite sign muons at
forward-rapidity (1.2~$<\mid$y$\mid<$~2.2), reconstructing the
invariant mass of the di-lepton pair, and subtracting the continuum
background~\cite{PHE07}.  The versatility of RHIC to provide heavy ion
collisions at different center of mass energies and for different ion
species has been further demonstrated during RUN-12 by providing first
\UU and \CuAu collisions at \snn = 193 and 200 GeV, respectively. Both
systems serve as an important test of the initial geometry.

\FigureThree
\subsection{\Jpsi in \AuAu collisions}
\label{subsec-4a}
The nuclear modification factor, \RAA, for \Jpsi as a function of
centrality (\Np) at mid-rapidity and forward-rapidity from \AuAu
collisions at \snn = 200 GeV is shown in Fig.~\ref{fig:AuAuRAA}. The
data show that the suppression of \Jpsi at forward-rapidity is
stronger than at mid-rapidity, even though the energy density is
expected to be slightly larger at mid-rapidity. This could be due to
stronger cold nuclear matter effects at forward-rapidity, or possibly
stronger coalescence of charm quarks at hadronization at mid-rapidity,
or both.

The comparison of the experimental data to the most recent theoretical
calculations that incorporate a variety of physics mechanisms
including gluon saturation, gluon shadowing, initial-state parton
energy loss, cold nuclear matter breakup, color screening, and charm
recombination are presented on Fig.~\ref{fig:AuAuRAA} and described in
details by PHENIX in Ref.~\cite{PHE11A}.
 
\subsection{\Jpsi in \CuAu collisions}
\label{subsec-4b}

Hot matter effects and CNM effects are present together in heavy ion
collisions, and both are important. In Au+Au collisions at RHIC, for
example, the addition of hot matter effects increases the suppression
of the \Jpsi by a factor of roughly two over what would be expected if
only CNM effects were present \cite{PHE11A,Bra2011}. By colliding ions
different sizes, like \CuAu, the cold nuclear matter effects is
expected to be different at forward and backward rapidities. The
comparison of d+Au, Au+Au and Cu+Au \Jpsi modifications across
rapidities may provide key insight on the balance of cold and hot
nuclear matter effects, and whether they are factorizable.

Figure \ref{fig:CuAuRAA} shows nuclear modification factor, \RAA, for
\Jpsi as a function of \Np for \CuAu collisions at \snn = 200 GeV
\cite{PHENIXCuAu}. The \RAA for \AuAu collisions at the same collision
energy and rapidity is shown in Fig. \ref{fig:CuAuRAA} for
comparison. The dependence of the \CuAu nuclear modification on \Np at
backward (Au-going) rapidity is similar to that for \AuAu collisions,
while the \CuAu \RAA at forward (Cu-going) rapidity is little lower.

\subsection{\Jpsi in \UU collisions}
\label{subsec-4c}
The study of collisions of deformed nuclei like $^{238}$U was
initially proposed \cite{Shu2000} because they promise an additional
gain of initial energy density relative to collisions of spherical
nuclei. A significantly deformed initial geometry at very high energy
density for specific orientations of \UU collisions is expected to
have observable effects on elliptic fow, jet quenching, \Jpsi
suppression and other observables that characterize the properties of
the quark gluon plasma (QGP) \cite{Shu2000,Heinz2005}. The interesting
configurations with high energy density can be selected in experiments
by using a combined cut on the elliptic flow value and the number of
spectators for high multiplicity events \cite{Kuhl2005}.

The nuclear modification factor, \RAA, for \Jpsi as a function of
centrality, \Np, at forward-rapidity (1.2~$<\mid$y$\mid<$~2.2) from
\UU collisions is shown in Fig.~\ref{fig:UU}. The \RAA for \AuAu and
\CuCu collisions at similar energy, \snn = 200 GeV, and rapidity are
also shown in Fig~\ref{fig:UU} for comparison. The centrality
dependence of \UU \RAA for \Jpsi shows a similar trend to the lighter
systems, \AuAu and \CuCu.

\section{Summary \label{sec-5}}
PHENIX has shown first measurements at RHIC from two very unique
systems: \UU and \CuAu at \snn = 193 and 200 GeV, respectively. Data
from both systems were compared to \AuAu and \CuCu collisions at
similar energy, \snn = 200 GeV. The comparison aimed to shed more
light on the role of the initial collision configuration and the final
data observables. The \UU system holds the highest energy density so
far measured at RHIC. We observed that the dependence of the \CuAu
nuclear modification for \Jpsi as function of centrality, \Np, at
backward (Au-going) rapidity is similar to that for \AuAu collisions,
while the \CuAu \RAA at forward (Cu-going) rapidity is noticeably
lower. Moreover, the centrality dependence of \UU \RAA for \Jpsi shows
a similar trend to the lighter systems, \AuAu and \CuCu at similar
energy. These observation lead to the conclusion that the centrality
dependence of \RAA for \Jpsi at given energy is less sensitive to the
system size.  Therefore these new PHENIX results of \RAA for \Jpsi in
\UU and \CuAu collisions are crucial for testing different theoretical
approaches.


%
%

\end{document}